%% file: fpcp11-rareB.tex
\documentclass[12pt]{article}
\usepackage{graphicx}
\usepackage{slashbox}
\usepackage{color}

\newcommand\pubdate{\today}

\newcommand\pubnumber{change/delete REPORT-\#}

\def\Title#1{\begin{center} {\Large #1 } \end{center}}
\def\Author#1{\begin{center}{ \sc #1} \end{center}}
\def\Address#1{\begin{center}{ \it #1} \end{center}}

\newcommand\pubblock{\rightline{\begin{tabular}{l} \pubnumber\\
         \pubdate  \end{tabular}}}
\newenvironment{Abstract}{\begin{center}{\bf Abstract}\end{center} \bigskip \begin{quotation}  }{\end{quotation}}
\newenvironment{Presented}{\begin{quotation} \begin{center} 
             PRESENTED AT\end{center}\bigskip 
      \begin{center}\begin{large}}{\end{large}\end{center} \end{quotation}}

\input econfmacros.tex
\def\mrm{\mathrm}
\def\ra{\rightarrow}

\newcommand{\afb}{\ensuremath{A_{FB}}}
\newcommand{\fl}{\ensuremath{F_{L}}}

\newcommand{\GeV}{\ensuremath{\mathrm{Ge\kern -0.1em V}}}
\newcommand{\MeV}{\ensuremath{\mathrm{Me\kern -0.1em V}}}

\newcommand{\pb}{\ensuremath{\mathrm{pb}^{-1}}}
\newcommand{\fb}{\ensuremath{\mathrm{fb}^{-1}}}

\newcommand{\bs}{\ensuremath{B^{0}_{s}}}
\newcommand{\bd}{\ensuremath{B^0_d}}
\newcommand{\bsd}{\ensuremath{B^{0}_{s,d}}}
\newcommand{\bu}{\ensuremath{B^+}}
\newcommand{\jp}{\ensuremath{J/\psi}}
\newcommand{\mm}{\ensuremath{\mu^{+}\mu^{-}}}
\newcommand{\mmk}{\ensuremath{\mu^{+}\mu^{-}K^{+}}}

\newcommand{\bmmks}{\ensuremath{\bd\ra\mm K^{*}(892)^0}}
\newcommand{\bpmmk}{\ensuremath{B^+\ra\mm K^+}}
\newcommand{\bmmphi}{\ensuremath{\bs\ra\mm \phi}}

\newcommand{\bsmm}{\ensuremath{\bs\ra\mm}}
\newcommand{\bdmm}{\ensuremath{\bd\ra\mm}}
\newcommand{\bsdmm}{\ensuremath{\bsd\ra\mm}}
\newcommand{\jpmm}{\ensuremath{\jp\ra\mm}}
\newcommand{\bjk}{\ensuremath{\bu\ra\jp K^{+}}}

\newcommand{\brbsmm}{\ensuremath{\mathcal{B}(\bsmm)}}
\newcommand{\brbdmm}{\ensuremath{\mathcal{B}(\bdmm)}}
\newcommand{\brbsdmm}{\ensuremath{\mathcal{B}(\bsdmm)}}

\newcommand{\Mmm}{\ensuremath{M_{\mm}}}

\newcommand{\pting}{\ensuremath{\Delta\alpha}}
\newcommand{\iso}{\ensuremath{Iso}}
\newcommand{\cdf}{CDF~II}
\newcommand{\dzero}{D0}

\begin{document}
\begin{titlepage}
\pubblock

\vfill

%%%%%%%%%%%%%%%%%%%%%%%%%%%%%%%%%%%%%%%%%%%%%%%%%%%%%%%
%%MODIFY
%%%%%%% TITLE, AUTHOR, ADDRESS 
%%%%%%%%%%%%%%%%%%%%%%%%%%%%%%%%%%%%%%%%%%%%%%%%%%%%%%%

\Title{Rare $B$ meson decays at the Tevatron}
\vfill
\Author{Walter Hopkins for the CDF and \dzero\ Collaborations}  
\Address{Cornell University, Ithaca, NY 14853}
\vfill

%%%%%%%%%%%%%%%%%%%%%%%%%%%%%%%%%%%%%%%%%%%%%%%%%%%%%%%
%%MODIFY
%%%%%%% Abstract
%%%%%%%%%%%%%%%%%%%%%%%%%%%%%%%%%%%%%%%%%%%%%%%%%%%%%%%

\begin{Abstract}
  Rare $B$ meson decays are an excellent probe for beyond the Standard Model physics. 
  Two very sensitive processes are the \bsdmm\ and $b\to s\mu^{+}\mu^{-}$ decays.
  We report recent results at a center of mass energy of $\sqrt{s} = 1.96$~TeV from the \cdf\ and \dzero\ collaborations
  using between 3.7 \fb\ and 6.9 \fb\ taken during Run II of the Fermilab Tevatron Collider. 

\end{Abstract}

\vfill

\begin{Presented}
The Ninth International Conference on\\
Flavor Physics and CP Violation\\
(FPCP 2011)\\
Maale Hachamisha, Israel,  May 23--27, 2011
\end{Presented}
\vfill

\end{titlepage}
\def\thefootnote{\fnsymbol{footnote}}
\setcounter{footnote}{0}
%

%%%%%%%%%%%%%%%%%%%%%%%%%%%%%%%%%%%%%%%%%%%%%%%%%%%%%%%
%%%%%%% Article body
%%%%%%%%%%%%%%%%%%%%%%%%%%%%%%%%%%%%%%%%%%%%%%%%%%%%%%%

\section{$b\to s\mu^{+}\mu^{-}$ decays}
$b\to s\mu^{+}\mu^{-}$ decays are flavor changing neutral current (FCNC) processes that can only occur through higher order box or penguin amplitudes
in the Standard Model. New physics can be probed by measuring various combinations of their decay rates. One of the most sensitive 
observables is the forward-backward asymmetry of the muons (\afb ) as a function of the squared di-muon mass. 

Three decays of interest are \bpmmk , \bmmks , and \bmmphi . 
The first two decays were observed at BaBar, Belle, and CDF. Belle found a 2.7$\sigma$ deviation for \afb\ for $B^{0}_{d}\rightarrow l^+l^- K^{*}$
\cite{belle}.
We describe hre the CDF analysis that uses 4.4 \fb\ of data~\cite{btosmmAsym}.

\subsection{Analysis Method}

\subsubsection{Branching Ratio Measurement}
Branching ratios for \bmmks, \bpmmk, and \bmmphi\ are 
measured relative to normalization modes, where the two muons originate from a $J/\psi$ decay. For the event
reconstruction CDF requires two muons with a transverse momentum ($p_T$) greater than either 1.5 GeV/c or 2.0 GeV/c depending on
the muon trigger. The three modes are then reconstructed where the $K^*(892)^0$ is reconstructed from $K^*(892)^0\to K^+\pi^-$ and the $\phi$ 
is reconstructed from $\phi\to K^+ K^-$. To avoid contamination from resonant modes
such as the $J/\psi$ and $\psi'$, candidates with di-muon masses near these resonances are rejected.

The events then have to meet loose preselection requirements before an artificial neural network (NN),
which combines multiple discriminating variables into one variable, is applied. Signal is modeled with $p_T$-reweighted 
Pythia signal Monte Carlo simulations (MC).
The reweighing is done by comparing the MC $p_T$ distribution with that of the normalization modes. The background is modeled by sampling 
the $B$ meson mass sideband regions. For the \bmmks\ and \bpmmk\ modes there are significant
physics backgrounds in the lower sideband region and thus only the higher sideband region between 5$\sigma$ and 15$\sigma$ ($\sigma$=20MeV/$\textnormal{c}^2$) above the known $B$ meson masses is used. For \bmmphi\ both lower and upper sidebands are used. 

The final signal yield is obtained by an unbinned maximum likelihood fit to the $B$ meson mass distribution. The probability distribution function (PDF) of the signal
is parametrized with two Gaussian with different means while the the background PDF is described by a first or second order polynomial. 
Peaking background contributions are subtracted from the fit results for the signal yields. The only significant peaking contribution is cross-talk among 
\bmmks\ and \bmmphi\ which has a $\sim$1\% contribution to the total observed signal MC yields. 
The final branching ratios are calculated as follows:

\begin{equation}\label{eq:brB}
  \mathcal{B}(B\to h\mu^{+}\mu^{-}) = \frac{N_{h\mu^{+}\mu^{-}}^{\mrm{NN}}}{N_{J/\psi h}^{\mrm{loose}}}
\cdot
\frac{\epsilon^{\mrm{loose}}_{J/\psi h}}{\epsilon^{\mrm{loose}}_{h\mu^{+}\mu^{-}}}
\frac{1}{\epsilon^{\mrm{NN}}_{h\mu^{+}\mu^{-}}}\cdot
  \mathcal{B}(J/\psi h) \cdot \mathcal{B}(J/\psi \to\mu^{+}\mu^{-} ) ,
\end{equation}

where $B$ signifies the $B^{0}_{d}$, $B^+$, or $B^{0}_{s}$ and $h$ represents $K^*(892)^0$, $K^+$, or $\phi$, 
$N_{h\mu^{+}\mu^{-}}^{\mrm{NN}}$ and $N_{J/\psi h}^{\mrm{loose}}$ are the yields after the optimal NN, 
$\frac{\epsilon^{\mrm{loose}}_{J/\psi h}}{\epsilon^{\mrm{loose}}_{h\mu^{+}\mu^{-}}}$ is the relative efficiency of the loose selection cuts, and 
$\epsilon^{\mrm{NN}}_{h\mu^{+}\mu^{-}}$ is the NN cut efficiency on the loosely-selected events. The NN is not applied to 
the normalization mode because the signal/purity and size is sufficient with the loose selection cuts.
The NN cut efficiency are obtained from signal MC.

The three leading systematics are the systematics on the efficiency, $\mathcal{B}(J/\psi h)$, and background PDF. 
The main sources within the efficiency systematic errors are the MC reweighing and the NN cut. The total 
systematics for the \bmmks, \bpmmk, and \bmmphi\ 
are 9\%, 7\%, and 32\%, respectively. 

\subsubsection{Forward-Backward Asymmetry Measurement}

For the \bmmks decay the forward-backward asymmetry (\afb ) as well as the $K^{*}$ longitudinal polarization (\fl)
are measured. These are extracted from $\cos\theta_{\mu}$, the cosine of the helicity angle between the $\mu^{+}$ ($\mu^-$) momentum vector and the opposite of $B$ ($\bar{B}$)
meson momentum vector in the di-muon rest frame, and $\cos\theta_K$, the cosine of the angle between kaon momentum and the opposite of the $B$ meson momentum vector in the $K^*(892)^0$ rest frame. B decay amplitudes are calculated using operator product expansion and Wilson coefficients.  
There are many non-SM predictions for \afb\ from models with different Wilson coefficients \cite{wilson}.

\afb\ and \fl\ are extracted using a unbinned maximum likelihood fit containing $B$ mass shape signal and background PDF's as well
as signal and background angular shape PDF's. The mass shape PDF's are divided into several di-muon mass bins and are described as 25 bin histograms.
The combinatorial background PDF is taken from the $B$ meson higher sideband. 
The angular acceptances are also described as 25 bin histograms and are derived from phase space signal MC. 

As a control \afb\ and \fl\ are fitted to $B^0_{d}\rightarrow J/\psi K^*(892)^0$ and \afb\ only to $B^+\rightarrow J/\psi K^+$. This cross check
yielded measurements that were consistent with other measurements. 

The systematic uncertainties for the \fl\ measurement in \bmmks\ are dominated by a fit bias ranging between 2\% and 7\%
for the different di-muon mass bins. This bias affects teh estimates of \afb\ or \fl\ when the true value
of these quantities are close to physical boundaries. The lowest di-muon mass bin also has a significant contribution 
from the angular efficiency systematic ($\sim$5\%) attained from changing the binning of the angular acceptance histograms. 
The total of all the systematics ranges between 2\% and 8\% for the different
di-muon mass bins. 

For the \afb\ measurement in \bmmks\ the dominant systematic uncertainties are the angular efficiency, fit bias, and the statistical
uncertainty from the \fl\ fit. 
The \fl\ fit systematic comes from varying \fl\ by $\pm 1\sigma$ where $\sigma$ is the statistical error from the \fl\ fit. The total
systematic errors range is 1\%-10\%.

The most significant systematic uncertainties in the \afb\ measurement for \bpmmk\ is given by the angular efficiency and angular background.
Angular background is estimated from the $B$ higher mass sideband. The systematic is estimated by changing the sideband region. The total
systematic uncertainty ranges between 3\% and 8\% for the different di-muon mass bins.

%The relevant Wilson coefficients for the $B^{0}\rightarrow \mu^+\mu^- K^{*}$ decay are $C_7$, 
%$C_9$, and $C_{10}$ which correspond to the electromagnetic penguin, vector electroweak, and axial-vector electroweak contributions, respectively. 

\subsection{Results}
\begin{figure}[htb]
  \centering
  \includegraphics[width=0.3\textwidth]{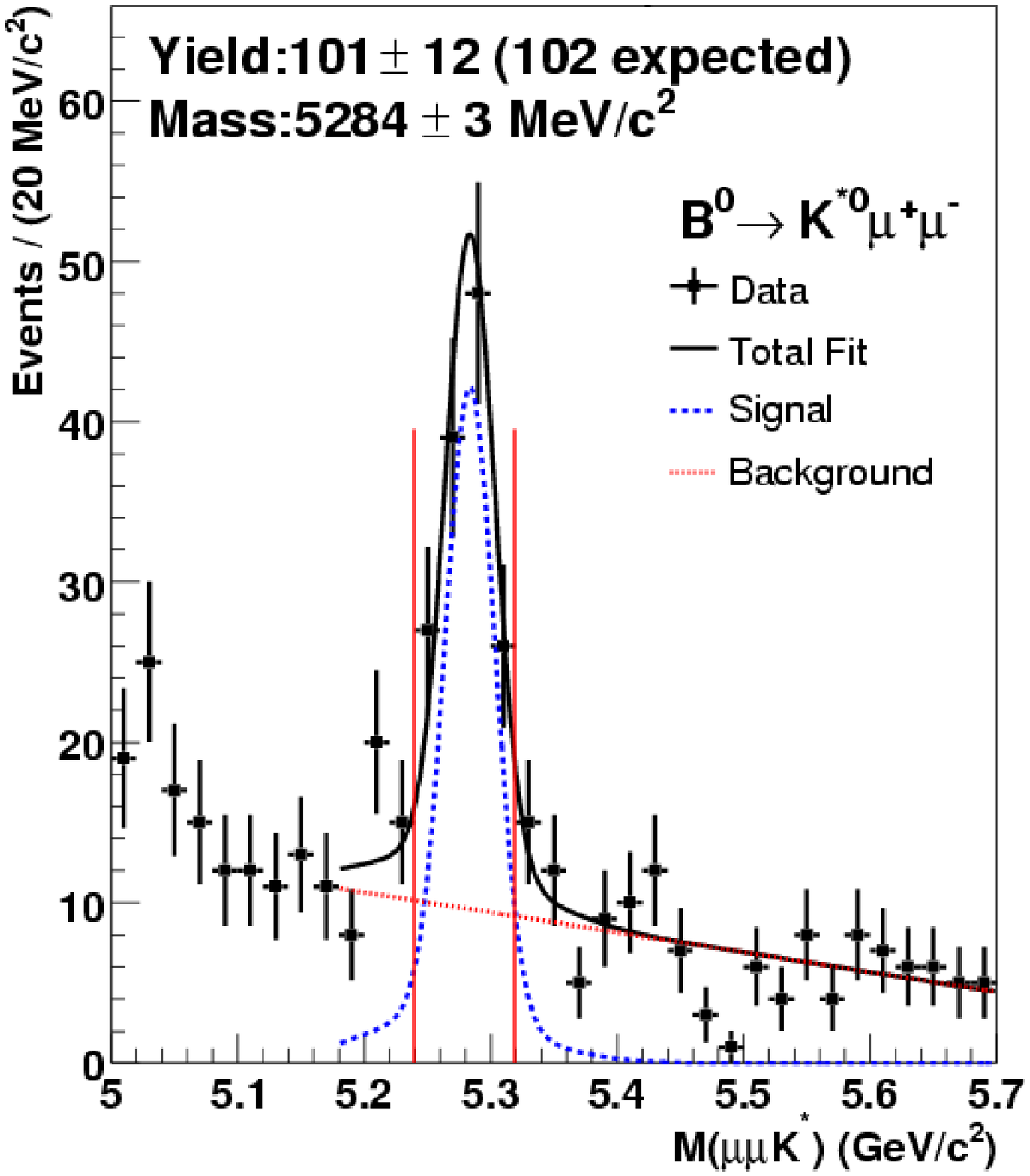}
  \includegraphics[width=0.3\textwidth]{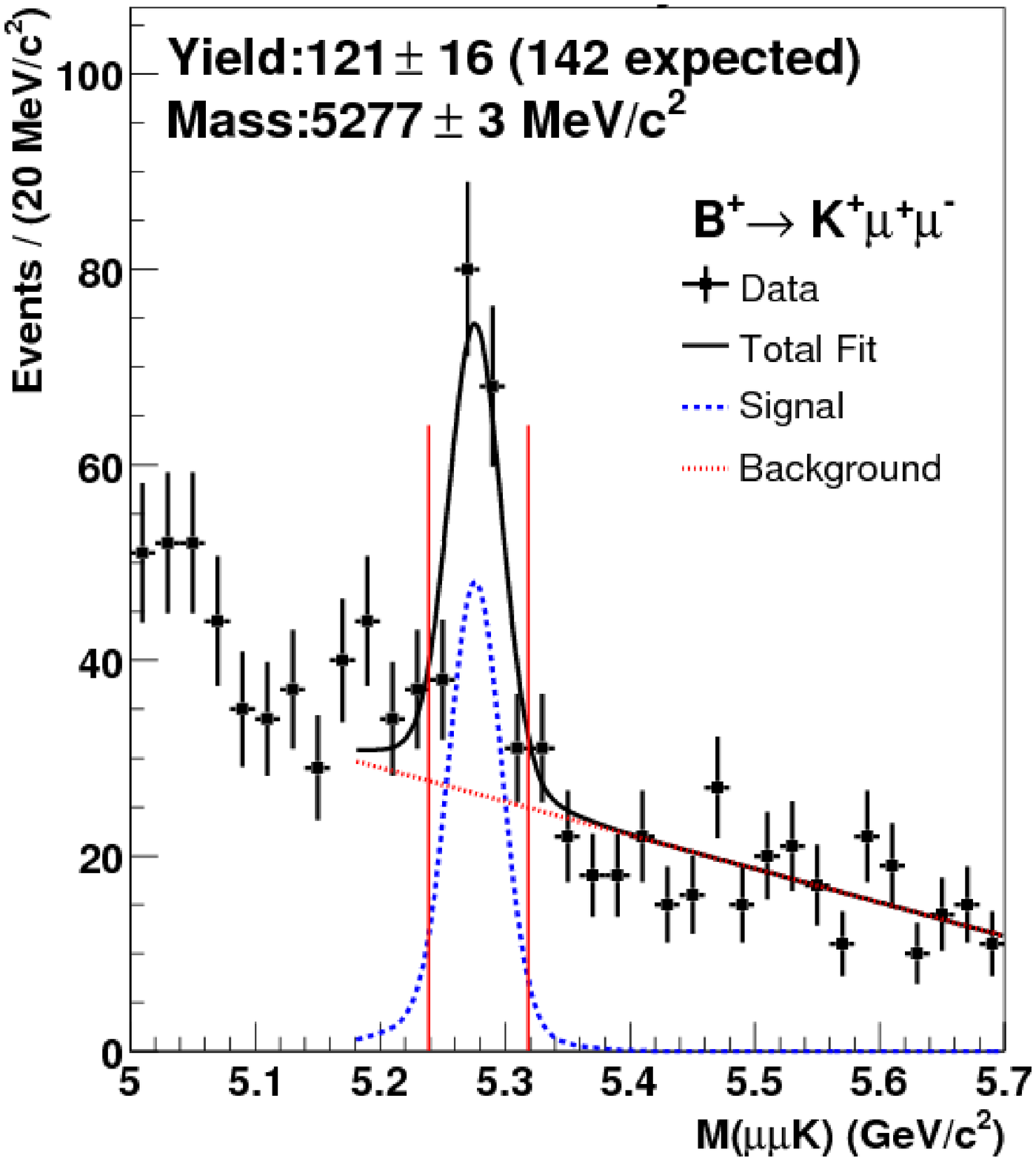}
  \includegraphics[width=0.3\textwidth]{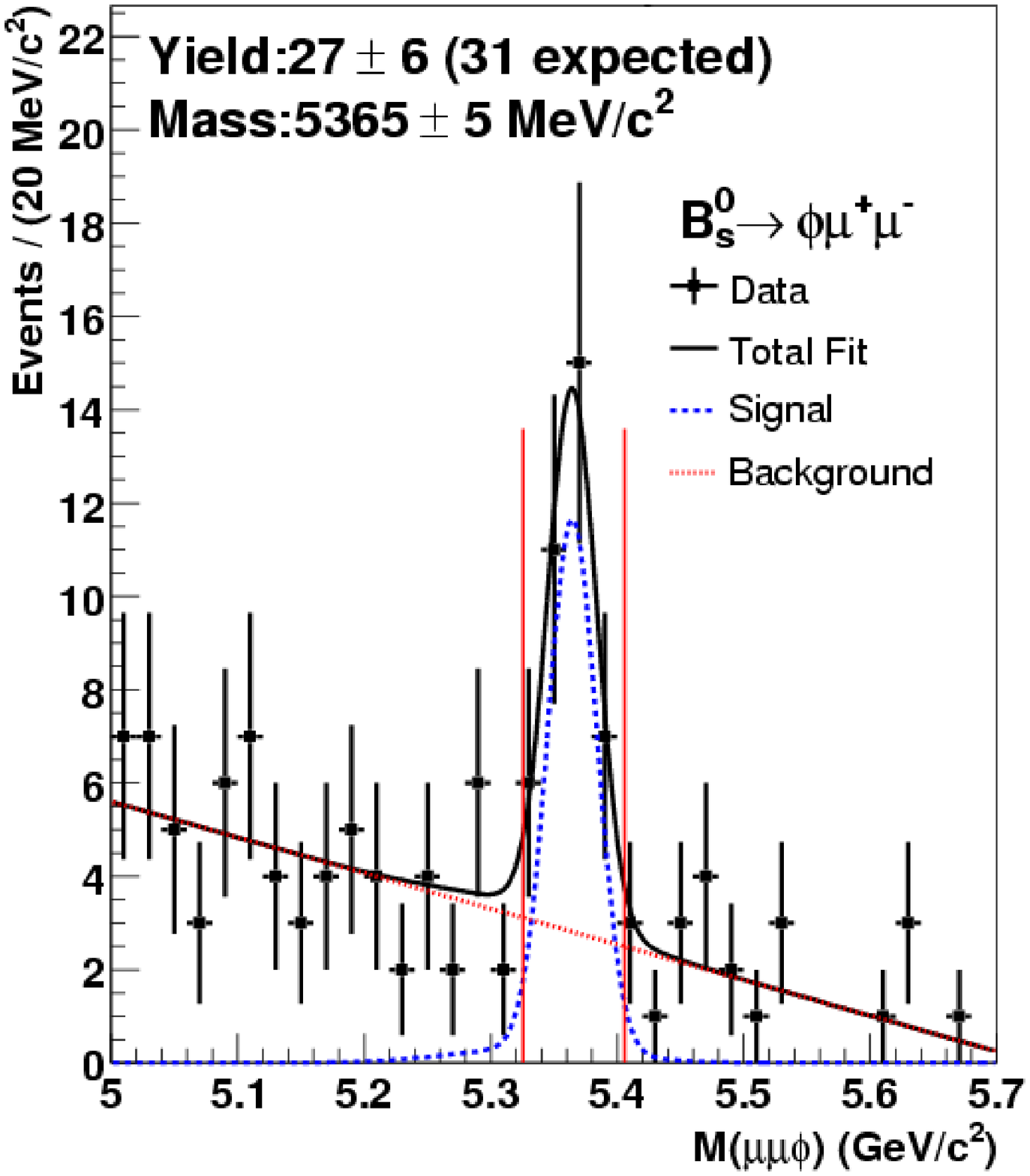}
  \caption{Yields for \bmmks, \bpmmk, and \bmmphi\ from left to right.
    In all histograms the red vertical lines indicate the signal regions.}
  \label{fig:btosmmYield}
\end{figure}

The resulting yield for the three decays are shown in Figure~\ref{fig:btosmmYield}. All measured branching ratios agree with previous measured 
values as well as theoretical predictions. CDF reports the first observation of \bmmphi with a significance of
$\sim 6\sigma$ and a measured branching ratio of  $\mathcal{B}(\bmmphi)=(1.44\pm 0.33$[stat]$\pm 0.46$[syst]$)\times 10^{-6}$.

\begin{figure}[htb]
  \centering
  \includegraphics[width=0.3\textwidth]{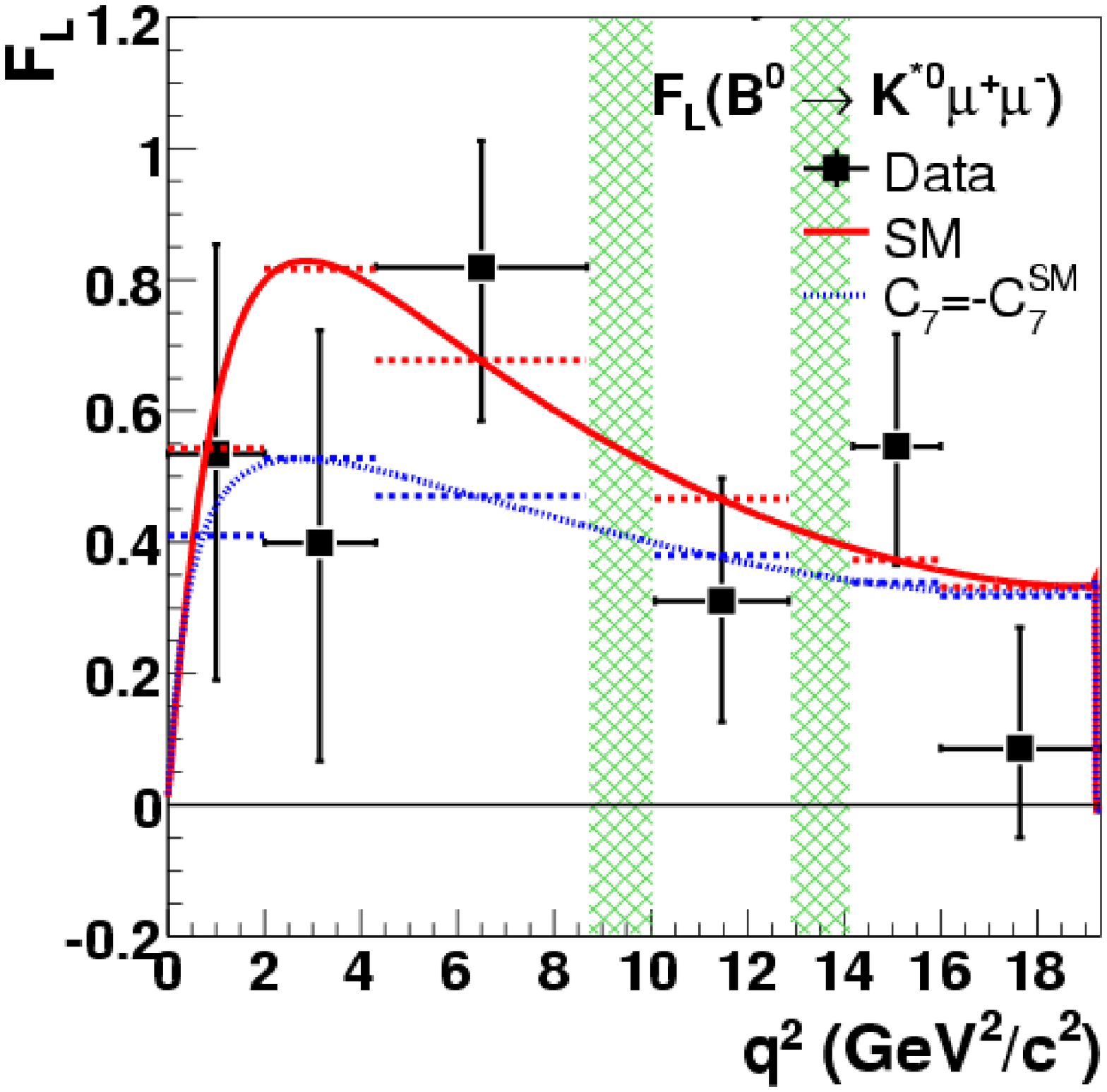}
  \includegraphics[width=0.3\textwidth]{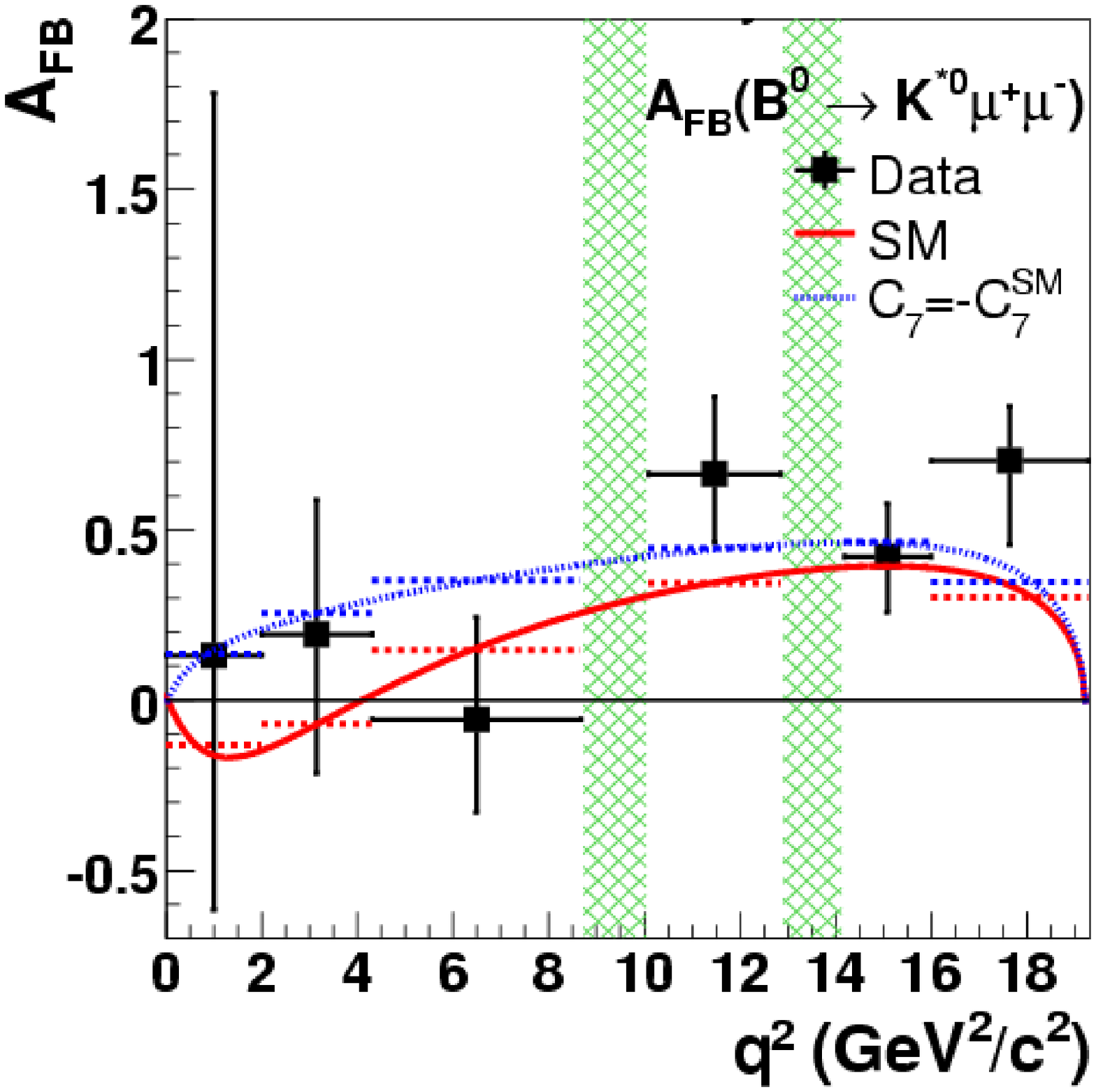}
  \includegraphics[width=0.3\textwidth]{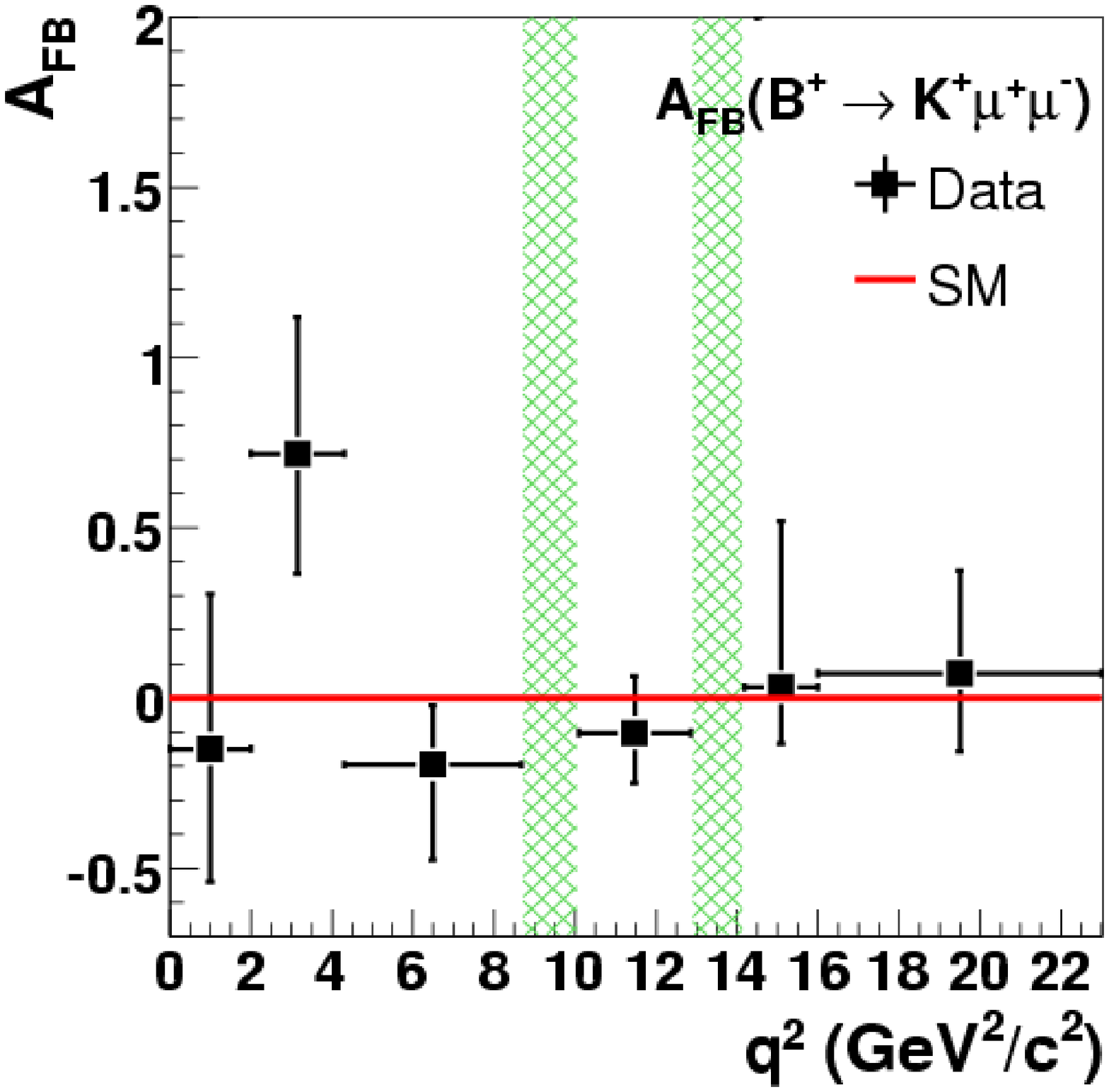}
  \caption{$K^{*}$ Longitudinal Polarization, \afb\ for \bmmks, and \afb\ for \bpmmk\ (from left to right).
The green hatched areas are due to the charmonium veto ($J/\psi$, $\psi'$).}
  \label{fig:afb}
\end{figure}

The results of the polarization and forward-backward asymmetry measurement shown in Figure~\ref{fig:afb} (\afb\ 
for \bpmmk\ is shown as a cross check) may contribute to contrain
several beyond SM physics models.  They are compatible and competitive with the results from the B-factories \cite{belle}, \cite{babar1},  \cite{babar2}.

\section{\bsdmm\ decays}

\bsdmm\ are FCNC decays that are highly suppressed by the SM.  The SM predictions for these branching fractions are $\brbsmm = (3.2\pm0.2)\times10^{-9}$ and $\brbdmm = (1.00\pm0.1)\times10^{-10}$~\cite{smbr}.
These predictions are one order of magnitude smaller than the current experimental sensitivity.   Previous bounds from the CDF collaboration, based on 3.7 $\:\fb$ of integrated luminosity, are $\brbsmm < 4.3\times10^{-8}$ and $\brbdmm < 7.6\times10^{-9}$ at $95\%$ C.L.~\cite{cdf9860}. \dzero\ bounds were previously set at 
 $\brbsmm < 5.1\times10^{-8}$ at $95\%$ C.L. using 6.1 \fb\ of data~\cite{d0bmm}. LHCb bounds of $\brbsmm < 5.6\times10^{-8}$ and $\brbsmm < 15\times10^{-9}$ at $95\%$ C.L were recently set using 37 \pb\ of data 
\vspace*{0.10in}

Enhancements to the rate of \bsmm\ decays occur in a variety of different new-physics models.  In supersymmetric (SUSY) models, new particles can increase \brbsmm\ by several orders of magnitude at large $\tan\beta$.  In the minimal supersymmetric standard model (MSSM), the enhancement is proportional to $\tan^6\!\beta$.  Global analysis including all existing experimental constraints suggest that the large $\tan\!\beta$ region is of interest~\cite{rparity,dark,chi2}.  For large $\tan\beta$, this search is one of the most sensitive probes of new physics available at the Tevatron experiments.  In contrast, SUSY R-parity violating models~\cite{rparity} and non-minimal flavor violating models~\cite{nmfv} can both enhance \bsmm\ and \bdmm\ separately even at low $\tan\!\beta$.  In the absence of an observation, limits on \brbsmm\ are complementary to those provided by other experimental measurements, and together would constrain the allowed supersymmetric parameter space.  
\vspace*{0.10in}

This document describes current status from \dzero\ and CDF, as well as the planned updated analysis by CDF. 

\subsection{Analysis Method}
\vspace*{0.10in}
Both CDF and \dzero\ collect opposite sign muon candidate events using di-muon triggers. 

\bjk\ events are collected on the same triggers as a relative 
normalization mode to estimate \brbsmm\ as:
\begin{equation}\label{eq:intro}
  \brbsmm = \frac{N_{\bs}}{\alpha_{\bs}\epsilon^{\mrm{total}}_{\bs}}\cdot
  \frac{\alpha_{B^{+}}\epsilon^{\mrm{total}}_{B^{+}}}{N_{B^{+}}}\cdot 
  \frac{f_{u}}{f_{s}}\cdot
  \mathcal{B}(\bjk)\cdot \mathcal{B}(\jpmm),
\end{equation}
where $N_{\bs}$ is the number of candidate \bsmm\ events, $\alpha_{\bs}$
is the geometric and kinematic acceptance of the di-muon trigger for
\bsmm\ decays, $\epsilon^{\mrm{total}}_{\bs}$ is the total efficiency
(including trigger, reconstruction and analysis requirements) for \bsmm\
events in the acceptance, with $N_{B^{+}}$, $\alpha_{B^{+}}$, and
$\epsilon^{\mrm{total}}_{B^{+}}$ similarly defined for \bjk\ decays.
The ratio $f_{u}/f_{s}$ accounts for the different $b$-quark fragmentation 
probabilities and is $(0.402\pm 0.013)/(0.112\pm 0.013) = 3.589 \pm 0.374$, 
including the (anti-)correlation between the uncertainties~\cite{PDG2010}. The final two terms are the relevant branching ratios 
$\mathcal{B}(\bjk)\cdot \mathcal{B}(\jpmm) = 
(1.01\pm0.03)\times10^{-3}\:\cdot\:(5.93\pm0.06)\times10^{-2}
=(6.01\pm0.21)\times10^{-5}$~\cite{PDG2010}.  
\vspace*{0.10in}

The CDF analysis described is also sensitive to \bdmm\ decays.  The value of \brbdmm\ 
is estimated from Equation~\ref{eq:intro} substituting \bd\ for \bs, and
changing $f_u / f_s$ to $f_u / f_d = 1$.  All other aspects are the
same as the \bsmm\ search except where noted below.
\vspace*{0.10in}
The analysis is done by first estimating the acceptances and efficiencies, then
creating a multivariate discriminant for background rejection. This discriminant is
optimized with Pythia signal MC and data mass sideband and validated with the \bjk\ normalization.
The background is then estimated, which has two sources: combinatorial background and peaking background (B$\rightarrow h^+h^-$).
Finally when the background is well understood the signal region is unblinded. 

\subsection{Signal and Background Properties}
\begin{figure}[htb]
  \centering
  \includegraphics[width=2.0in]{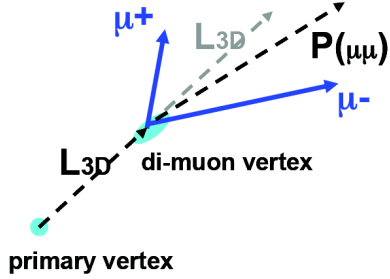}
  \caption{Sketch of the decay topology.}
  \label{fig:sigDecayVtx}
\end{figure}

The signal candidates are fully reconstructed events with a secondary vertex due to the long lifetime ($\sim450\mu$m) of the \bs\ meson. 
Signal events will have a primary-to-secondary vertex vector that is aligned with the \bs\ candidate momentum vector (Figure~\ref{fig:sigDecayVtx}).
Another property of signal events that is unique is that they are very isolated (with few tracks near the muon tracks) due to the hard B fragmentation.

Background events tend to be partially reconstructed and be shorter lived than signal. 
They also have a softer $p_T$ spectrum, higher activity of tracks, and misaligned primary-to-secondary vertex and momentum vectors. 
The combinatorial background consists of sequential semi-leptonic decay ($b\rightarrow c\mu^- X \rightarrow \mu^+\mu^-X$) and 
and double semi-leptonic decay ($bb\rightarrow\mu^-\mu^+X$). 
The expected number of combinatorial background events in the signal window 
is estimated by extrapolating the number of events in the sideband regions 
to the signal window using a first order polynomial fit. 

The peaking background from two-body hadronic $B$ decays is also evaluated.
These backgrounds are about a factor ten smaller than the combinatorial background and  need to be estimated separately.
The dominant sources of background are the decays of $B^{0}_{s}$ and $B^0$ to $h^+ h^{\prime -}$ final
states, where $h$ or $h^\prime$ can be either $\pi^\pm$ or $K^\pm$ and are misreconstructed (fake muons).
% ==============================================================

% ===========================================================
\subsection{CDF Analysis with 3.7 \fb\ of Data}

\subsubsection{Signal Discrimination}
\begin{figure}[htb]
  \centering
  \includegraphics[width=0.45\textwidth]{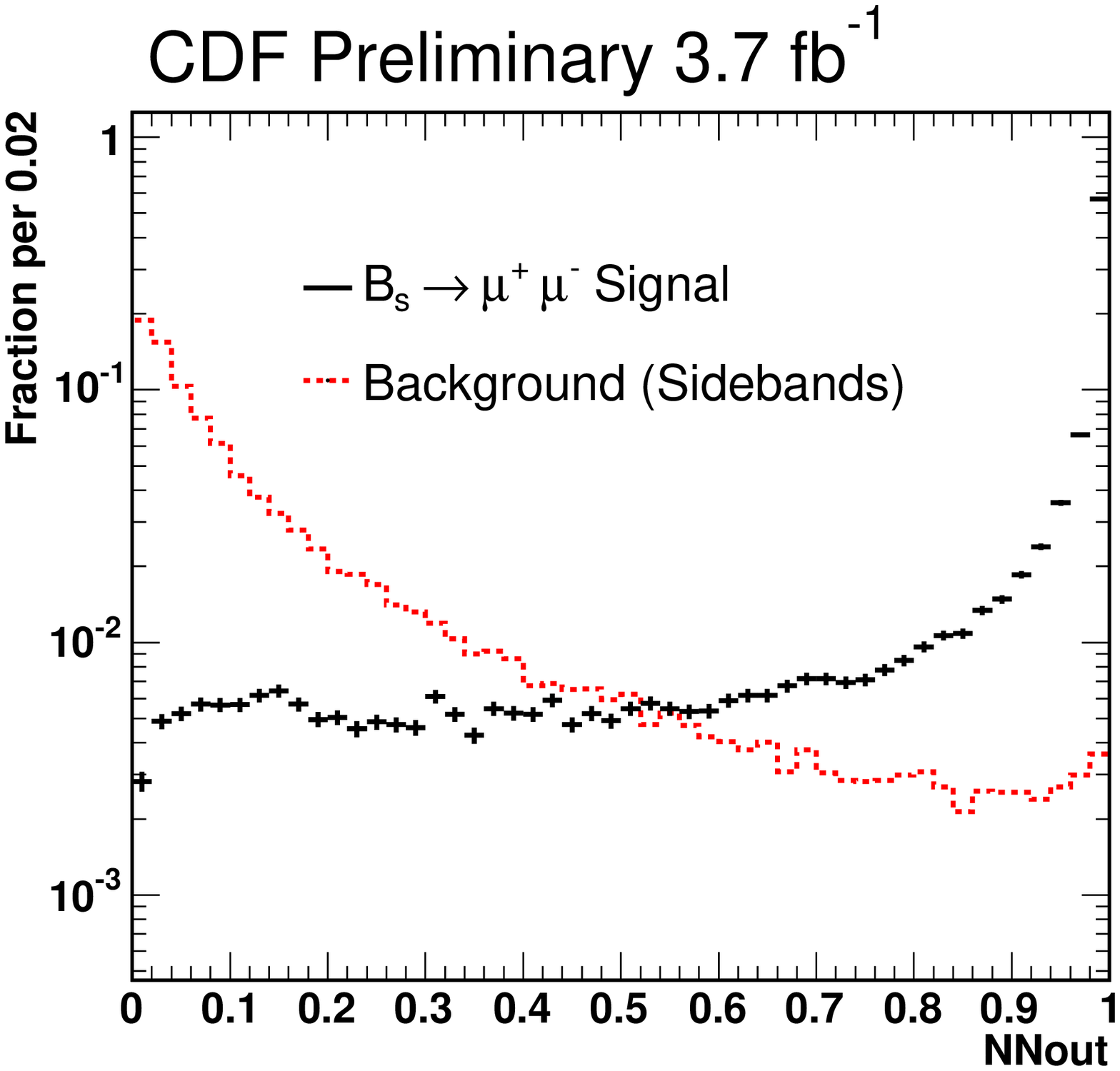} 
  \includegraphics[width=0.45\textwidth]{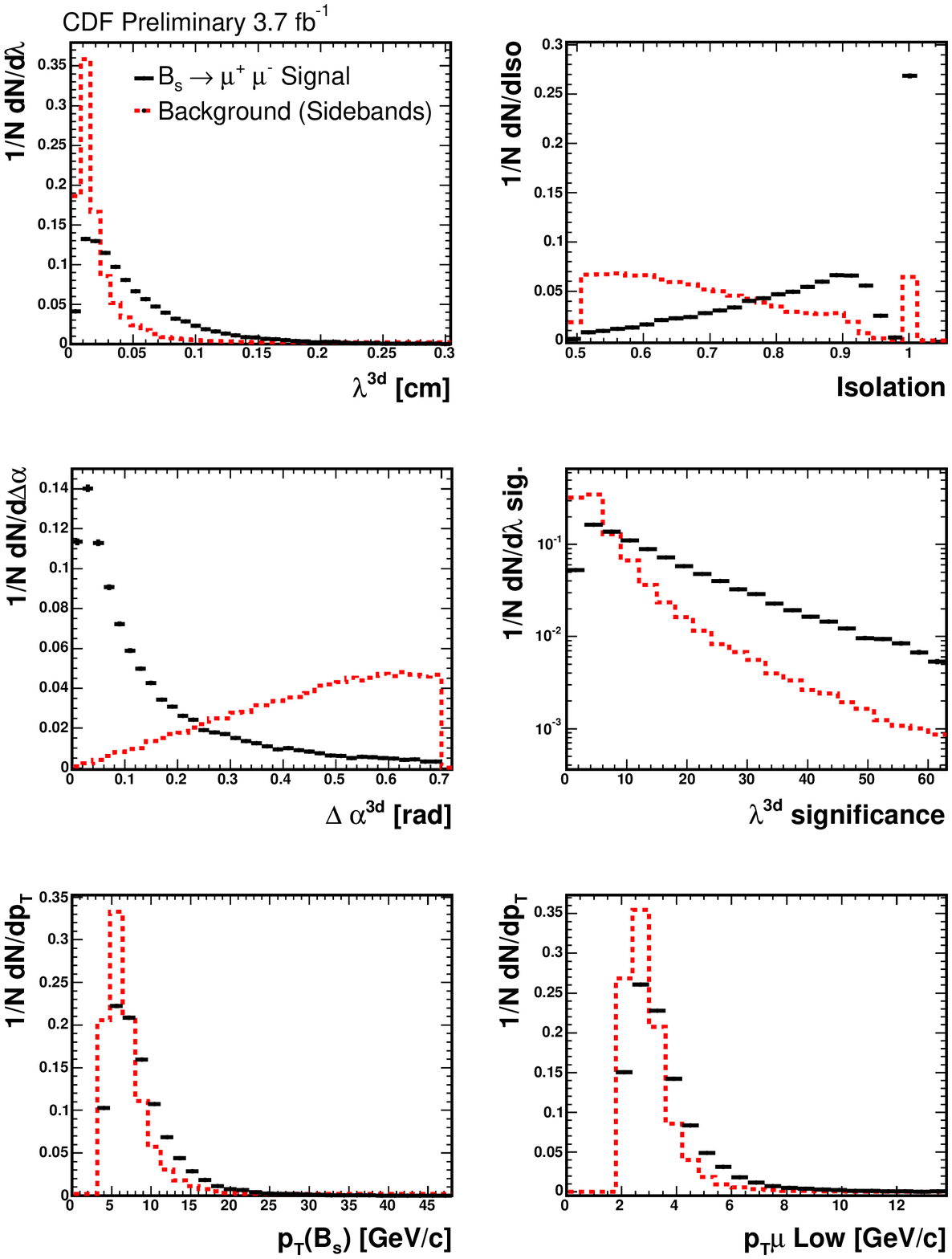}
  \caption{\textbf{Left:} Neural network output for signal MC and data mass sideband.
  \textbf{Right}: Distributions of neural network input variable for signal MC and data mass sideband.}
  \label{fig:nnInOut}
\end{figure}

Six variables are used for the signal discrimination (Figure~\ref{fig:nnInOut}). Three of the variables are secondary vertex variables.
These variables are combined in an artificial neural network which is optimized with signal MC
and data mass sidebands (Figure~\ref{fig:nnInOut}, left). 
The NN output is then divided into three bins while the di-muon mass is split into 5 
mass bins for limit calculation.

\begin{description}
  \item[$\mathbf{\lambda}$:] the proper decay length;

  \item[$\mathbf{\Delta\alpha}$:] the 3D opening angle between the \bs\ flight 
    direction, $\vec{p}(B)$, and the direction of the decay vertex - estimated
    as the vector originating at the primary vertex and terminating at the 
    muon-pair vertex;

  \item[$\mathbf{\iso}$:] the isolation of the candidate \bs\ defined as
    $\iso=p_{T}^{\bs}/(p_{T}^{\bs}+\sum_i p_{T}^{i}(\Delta R<1.0))$,
    where the sum is over all tracks within an $\eta-\phi$ cone of radius
    $R=1.0$, centered on $\vec{p}(B)$;

  \item[$\mathbf{\lambda}\mathbf{/}\mathbf{\sigma_{\lambda}}$:] proper decay length significance 
        ($\lambda$ divided by the uncertainty on $\lambda$);

  \item[$\mathbf{p_T(B^{0}_{s})}$:] the transverse momentum of the $B^{0}_{s}$ candidate;

  \item[$\mathbf{p_T (\mu)}$:] the transverse momentum of lower momentum
  daughter muon.		
\end{description}

\subsubsection{Control Samples for Combinatorial Background Estimates}
\label{sec:control}

CDF cross checked the background estimate procedure on four independent background samples:
\begin{description}
\item[OS-:]  opposite-sign muon pairs, which pass the baseline and vertex cuts
  and have a negative lifetime;
\item[SS+:]  same-sign muon pairs, which pass looser baseline and vertex cuts
  and have a positive lifetime;
\item[SS-:] same-sign muon pairs, which pass looser baseline and vertex cuts 
  and have a negative lifetime;
\item[FM+-:]  opposite-sign fake-muon pairs, at least one leg of which is 
  required to {\it{fail}} the muon ID requirement, passing 
  looser baseline and vertex cuts for both positive and negative lifetimes.
\end{description}

\vspace*{0.10in}
These are representative of various background contributions and thus are excellent control regions
to test the method of background estimation.
We compare the estimated background with actual observation. 
The result of this cross check is summarized in Table~\ref{tab:Xchecks}.
The agreement between predicted and observed number of background events in the signal region is good across all of
our control regions. The backgrounds were estimated using the same method as for the signal sample.

\begin{table}[!hbtp]
\begin{center}
\begin{tabular}{|cc||ccc|ccc|} \hline
          & & \multicolumn{3}{c|}{CMU-CMU}  
            & \multicolumn{3}{c|}{CMU-CMX}  \\
    sample  & $NN$ cut  & pred          & obsv  & prob(\%) & pred   &obsv & prob(\%)\\ 
\hline\hline
            & $0.80<\nu_{NN}<0.95$ & $275\pm (9)$   & $287$ & 26& $310\pm(10)$   & $304$& 39\\
     OS-    & $0.95<\nu_{NN}<0.995$ & $122\pm (6)$  & $121$ & 46& $124\pm(6)$   & $148$& 3.2\\
            & $0.995<\nu_{NN}<1.0$ & $44\pm(4)$    & $41$  & 36& \color{red}{$31\pm(3)$}    & \color{red}{$50$} & \color{red}{0.4}\\ \hline
            & $0.80<\nu_{NN}<0.95$ & $2.7\pm(0.9)$  & $1$   & 29& $2.7\pm(0.9)$ & $0$  & 10\\
     SS+    & $0.95<\nu_{NN}<0.995$ & $1.2\pm(0.6)$  & $0$   & 34& $1.2\pm(0.6)$ & $1$  & 66\\
            & $0.995<\nu_{NN}<1.0$ & $0.6\pm(0.4)$ & $0$   & 55& $0.0\pm(0.0)$ & $0$  & - \\ \hline
            & $0.80<\nu_{NN}<0.95$ & $8.7\pm(1.6)$  & $9$    & 49& $5.7\pm(1.6)$ & $2$  & 11\\
     SS-    & $0.95<\nu_{NN}<0.995$ & $3.0\pm(1.0)$  & $4$  & 36& $3.6\pm(1.0)$ & $2$  & 34\\
            & $0.995<\nu_{NN}<1.0$ & $0.9\pm(0.5)$  & $0$ & 43& $0.3\pm(0.3)$ & $0$  & 70 \\ \hline
            & $0.80<\nu_{NN}<0.95$ & \color{blue}{$169\pm (7)$} & \color{blue}{$169$} & \color{blue}{50}& $73\pm(5)$ & $64$  & 19 \\
     FM+    & $0.95<\nu_{NN}<0.995$ & $55\pm (4)$ & $43$  & 9& $19\pm(2)$ & $18$  & 49 \\
            & $0.995<\nu_{NN}<1.0$ & $20\pm(2)$ & $20$  & 48& $3.6\pm(1.0)$ & $3$  & 53\\ \hline
\end{tabular}
\caption{ 
  The values given in the parentheses are the uncertainties on the mean of the
  background prediction.
  The Poisson probability for making an observation at least as large (or fewer
  than observed when observed is less than predicted) given the 
  predicted background is also shown in the table. The best (blue) and worst (red) cases are also shown.}
\label{tab:Xchecks}
\end{center}
\end{table}

\vspace*{0.10in}

\subsubsection{Results}
\begin{figure}[htb]
  \centering
  \includegraphics[width=3.0in]{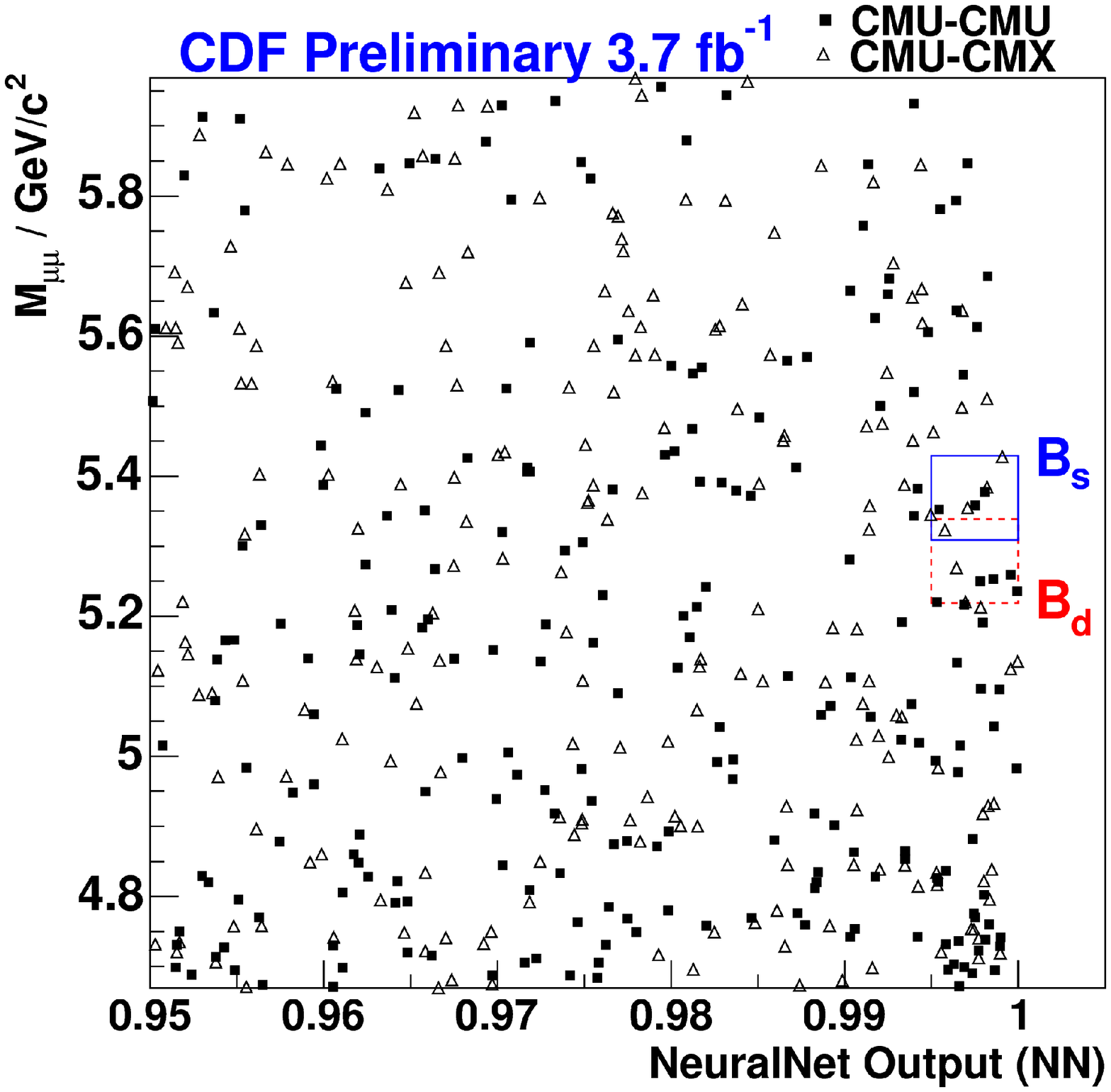}
 \caption{Di-muon mass vs NN output distribution for central-central (CMU-CMU) and central-forward (CMU-CMX) muons for CDF analysis.}
  \label{fig:2dPlot}
\end{figure}

The results from the CDF analysis on 3.7 \fb\ are shown in Table~\ref{tab:resultsCDF} and Figure~\ref{fig:2dPlot}.
$6.1\pm1.8$ events were expected in the signal region and 7 events were observed, resulting in the observed limits, using 3 NN bins and 5 mass bins, shown in Table~\ref{tab:resultsCDF}.

\begin{table}[!hbtp]
\begin{center}
\begin{tabular}{|l|cc|}
\hline \textbf{Channel}& \textbf{Expected} & \textbf{Observed} \\ 
\hline \bs\ Central &  4.0$\pm$1.0 & 3\\
\hline \bs\ Extended & 2.1$\pm$0.8 & 4\\
\hline \bd\ Central &  5.3$\pm$1.0 & 5\\
\hline \bd\ Extended & 2.8$\pm$0.8 & 3\\
\hline
\end{tabular}
\vspace*{0.10in}

\begin{tabular}{|l|cc|}
\hline& \textbf{90\% CL} & \textbf{95\% CL} \\ 
\hline \bs\ &  3.6$\times 10^{-8}$ & 4.3$\times 10^{-8}$\\
\hline \bd\ &  6.0$\times 10^{-9}$ & 7.6$\times 10^{-9}$\\
\hline
\end{tabular}

\caption{\bs\ and \bd\ expected and observed number of candidates in central and extended mass region (top) as well as resulting observed
limits (bottom).}
  \label{tab:resultsCDF}
\end{center}
\end{table}

\subsection{\dzero\ Analysis with 6.1 \fb\ of Data}

\begin{figure}[htb]
  \centering
  \includegraphics[width=3.5in]{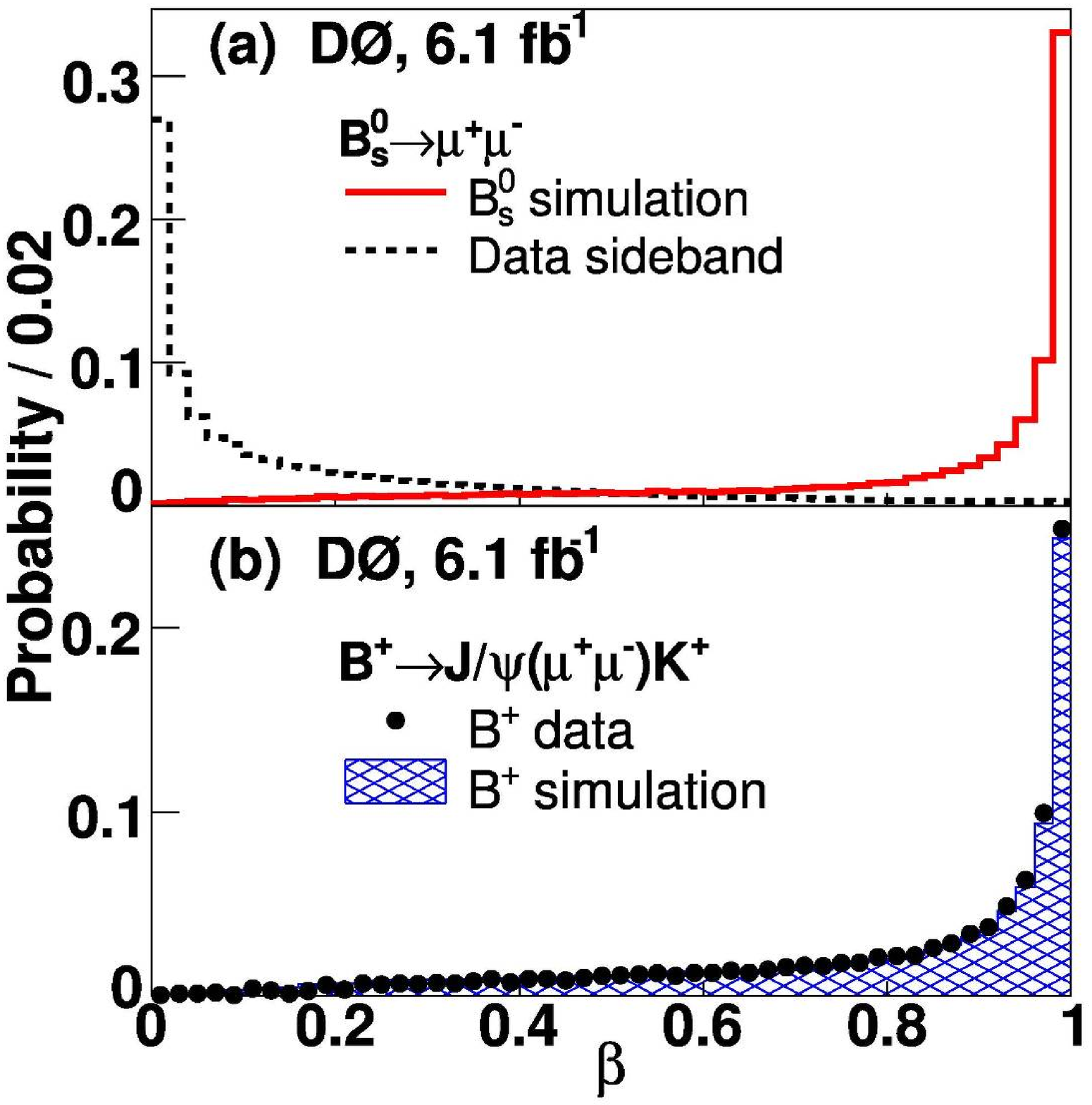}

  \caption{Distribution of Bayesian neural network output for signal MC and data mass sideband for \dzero\ analysis.}
  \label{fig:d0NN}
\end{figure}

\dzero\ uses a similar method as described above. For signal discrimination, variables are combined into a Bayesian Neural Network that also
uses signal MC and data mass sidebands for optimization (Figure~\ref{fig:d0NN}). Unlike CDF, \dzero\ does not have the mass resolution to distinguish
between \bd\ and \bs .

\begin{figure}[htb]
  \centering
  \includegraphics[width=3.0in]{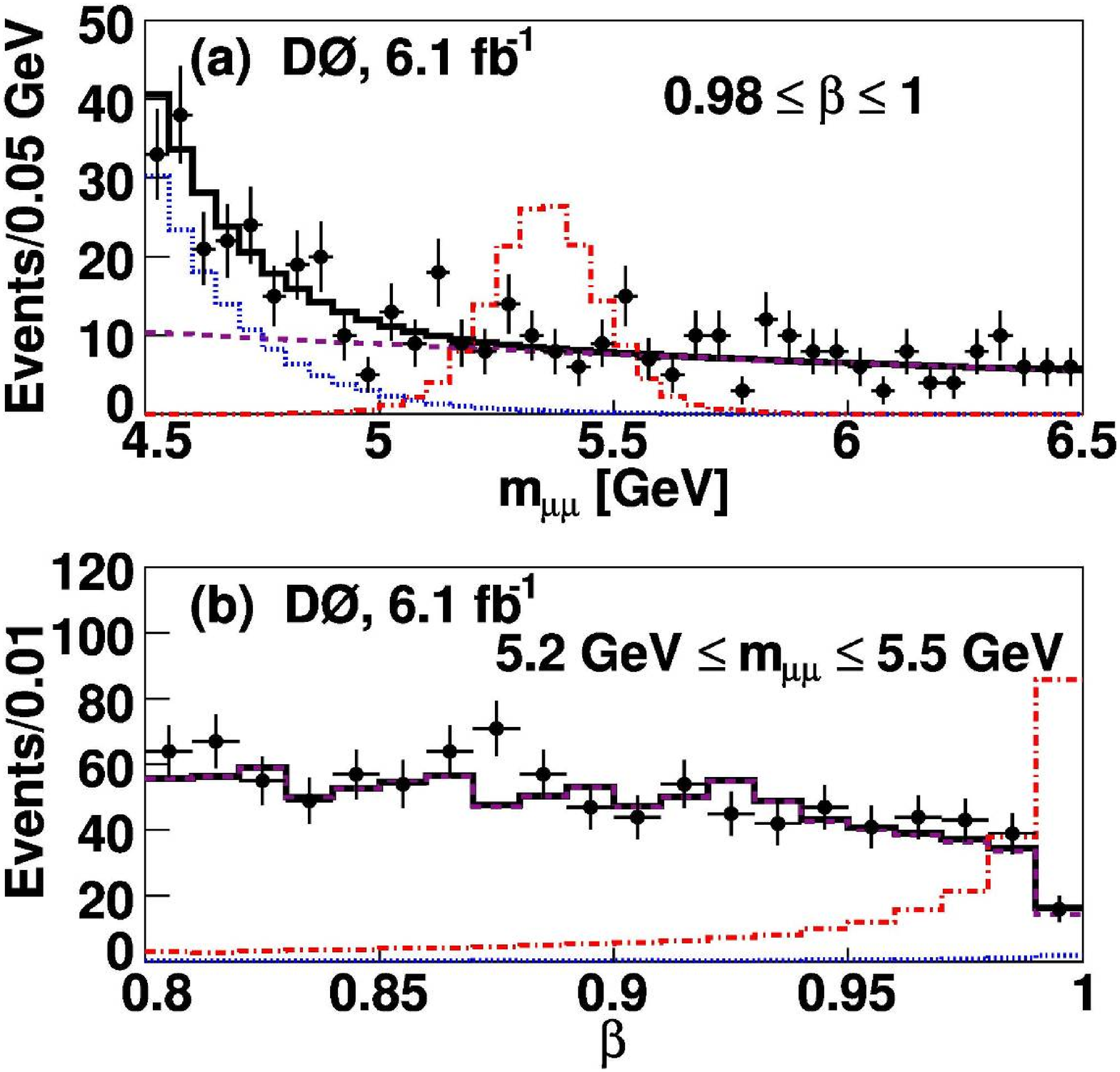}

  \caption{\dzero\ unblinded di-muon mass distributions. The dots with error are the data, the solid black line
    is the expected number of background events, the red dotted-dashed lines is the SM signal $\times 100$, the
    dashed line is $B(D)\rightarrow\mu^{+}\nu X$ and $\bar{B}(\bar{D})\rightarrow\mu^{-}\bar{\nu} X'$, and the blue dotted line
    is $B\rightarrow\mu^{+}\nu \bar{D}$ and $\bar{D}\rightarrow\mu^{-}\bar{\nu}X$}.
  \label{fig:d0Result}
\end{figure}

The \dzero\ results are shown in Figure~\ref{fig:d0Result}. 51$\pm$4 background events are expected
in the highest sensitivity region and 55 events are observed. The expected $B^{0}_{s}$ limit is $3.8 \times 10^{-8}$ at 95\% CL
observed a limit of 5.1$\times 10^{-8}$.

\subsection{CDF Improvements}
\begin{figure}[htb]
  \centering
  \includegraphics[height=3.5in]{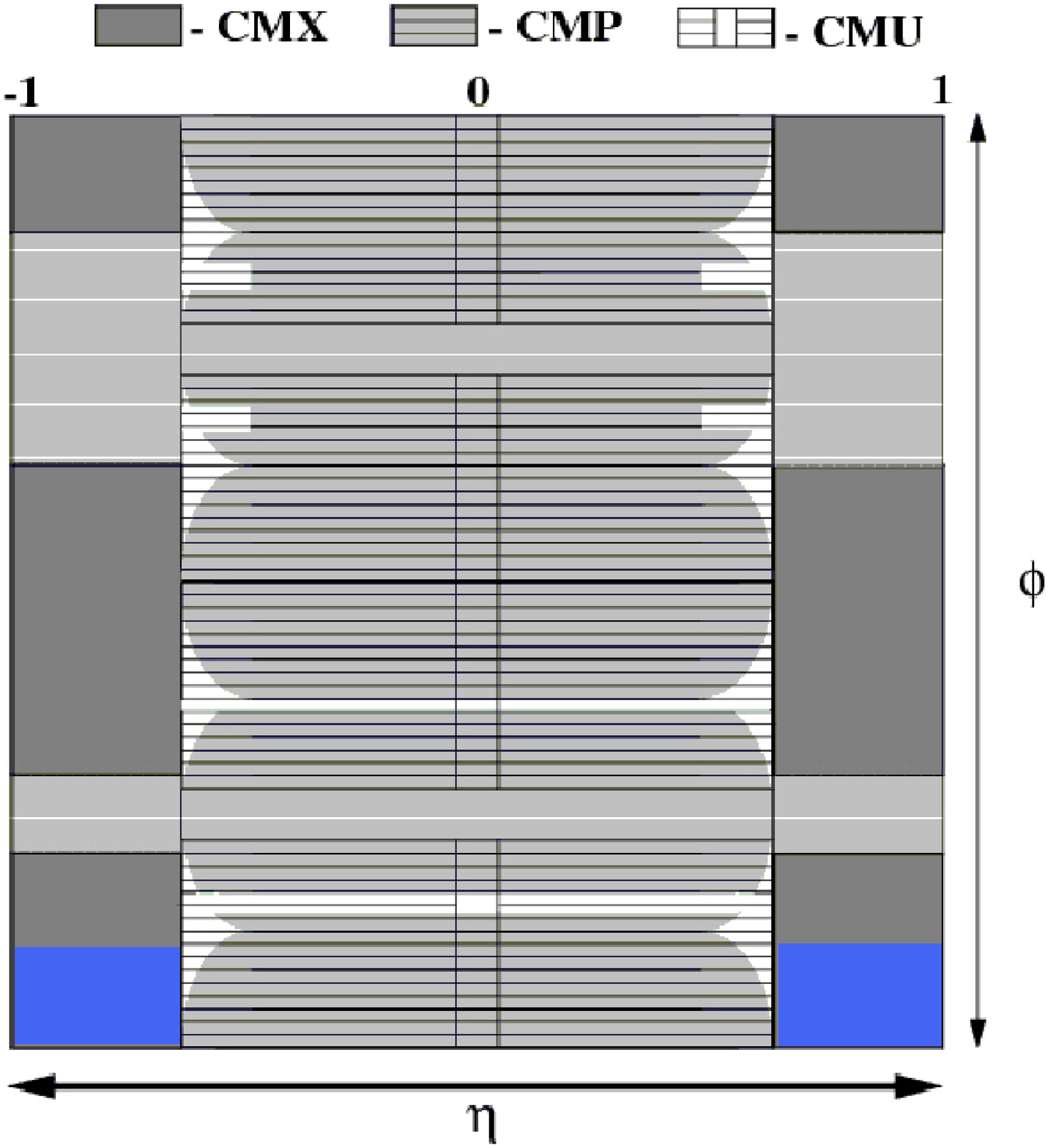}
  \includegraphics[height=3.5in]{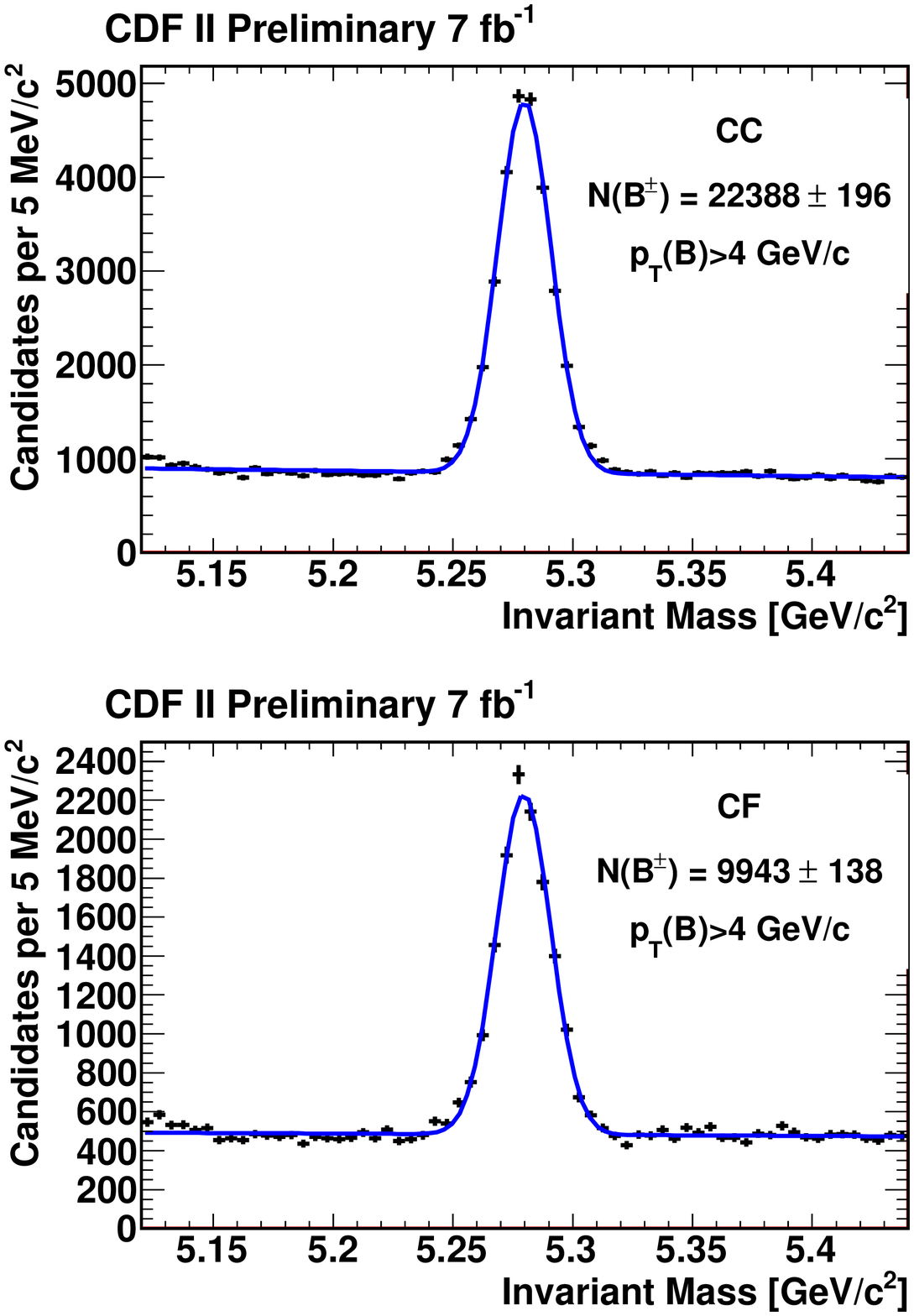}
  \caption{\textbf{Left:} CDF Muon detector layout. The blue region has been added for the updated analysis.
\textbf{Right:} The \mmk\ invariant mass distribution for events
    satisfying the baseline and vertex requirements for the \bjk\ sample.}
  \label{fig:miniSkirtsBpYields}
\end{figure}

\begin{figure}[htb]
  \centering
  \includegraphics[width=0.4\textwidth]{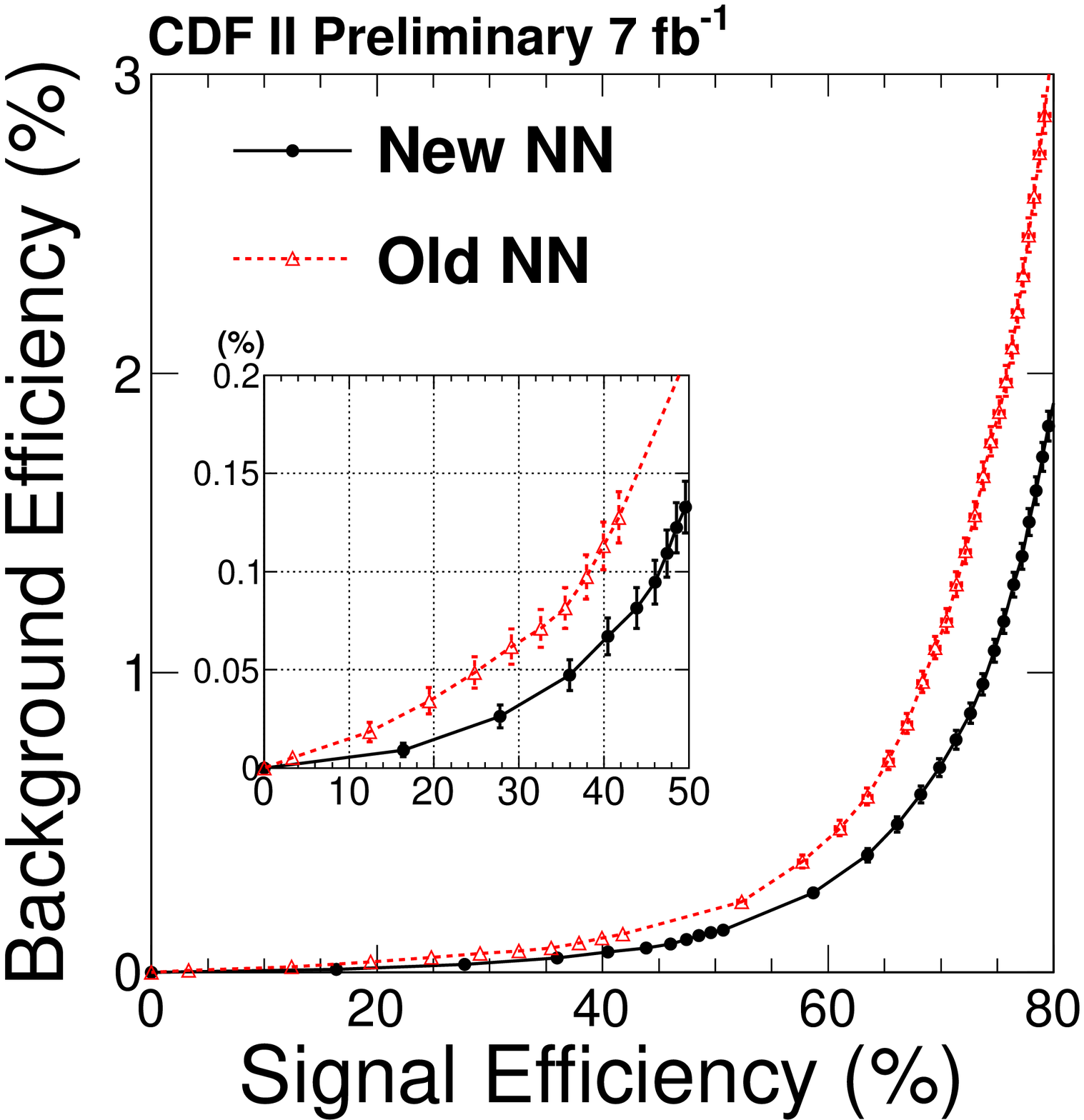}
  \includegraphics[width=0.43\textwidth]{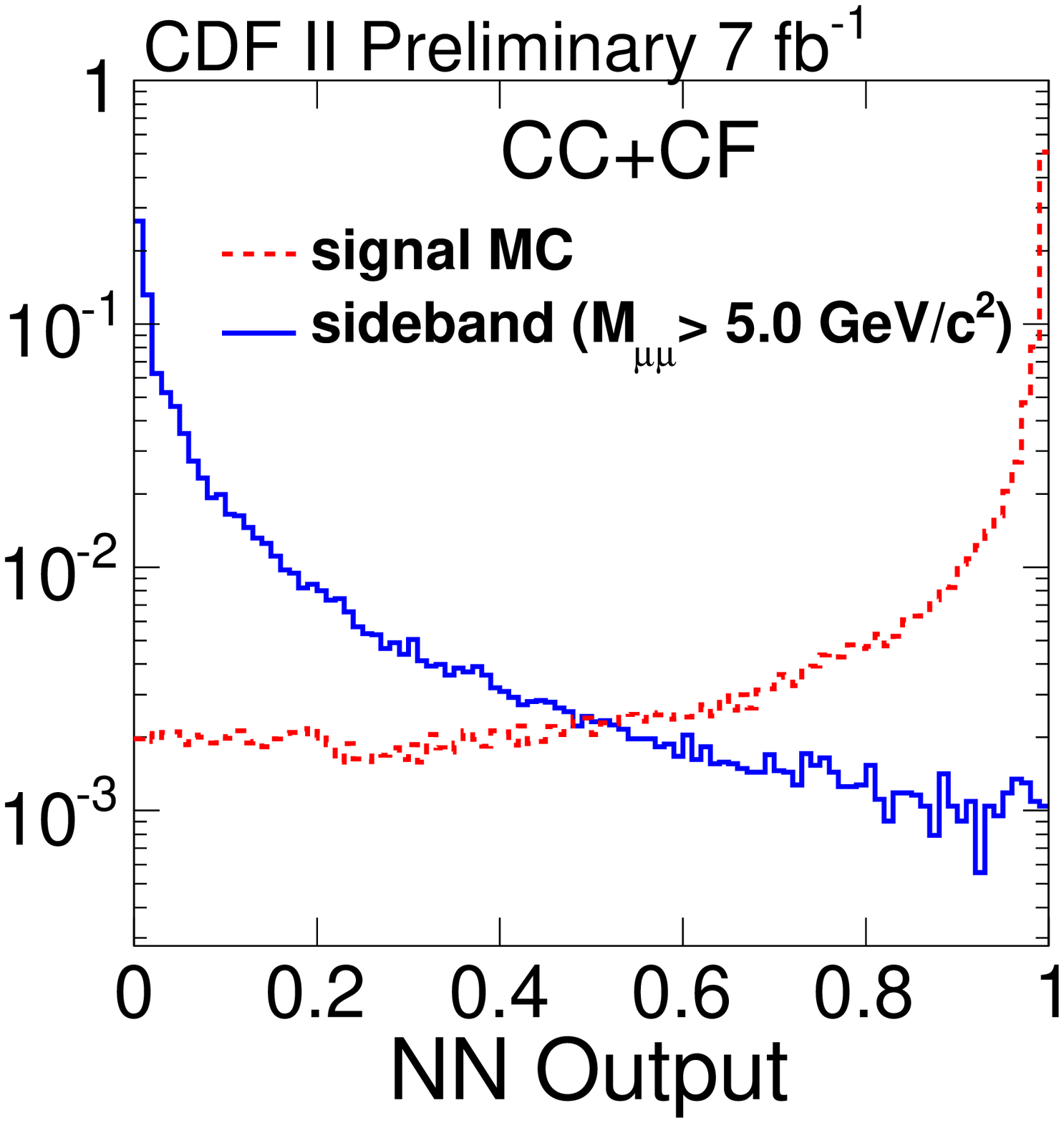}
  \caption{\label{fig:NNoutEff}\textbf{Left:} Signal and background efficiency for both the old and new neural network. 
    \textbf{Right:} The distribution of NN output for CC and CF combined.  
    The red histogram is the signal MC, the black histogram is sideband data.}
\end{figure}

\begin{figure}[htb]
  \centering
  \vspace*{-0.10in}
  \includegraphics[width=0.6\textwidth]{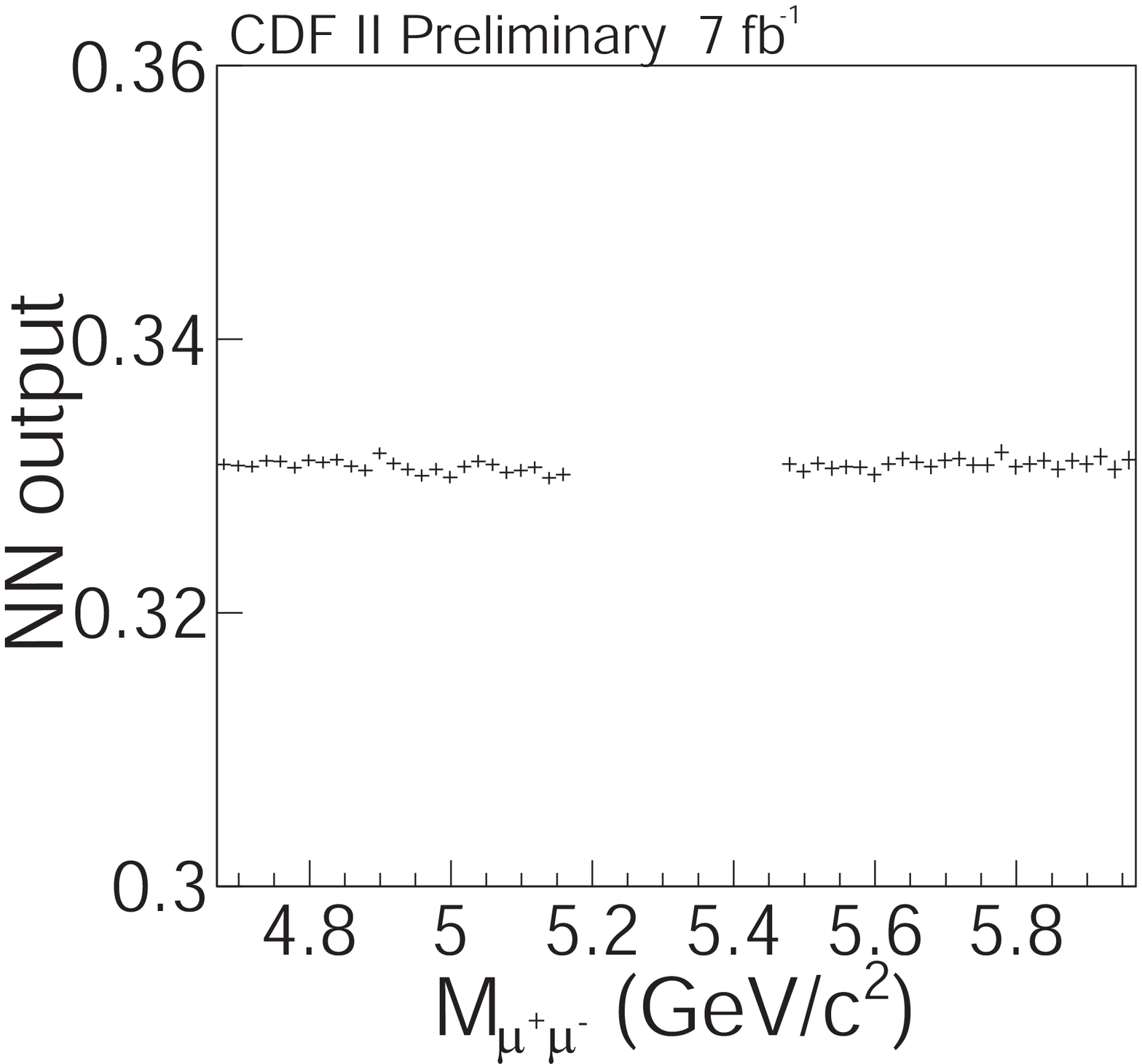}
  \caption{\label{fig:backsculpt} 
    The top plot shows the NN output distributions for background (black histogram) and background
    trained as ``signal'' (red histogram).  The bottom plot shows NN output as a function of di-muon invariant mass.}
\end{figure}

CDF is working on an improved analysis with significantly more (3.2 \fb) data. The forward muon acceptance (left of Figure~\ref{fig:miniSkirtsBpYields}) 
has been increased and the neural network has been improved to achieve greater background rejection. The peaking background prediction has
also been improved. The central-forward (CMU-CMX) channel of the analysis has increased in statistics by $\sim$15\% resulting
in a total increase for both channels of $\sim$7\%.
The neural network has been improved by increasing the background rejections (Figure~\ref{fig:NNoutEff}). Additionally
the neural network has been extensively tested for mass bias (Figure~\ref{fig:backsculpt}).

The following variables, shown in order of significance to NN separation power, are used to discriminate the \bsmm\ signal from
the background:

\begin{description}

\item[$\mathbf{\pting}$:] the 3D opening angle between the \bs\ flight 
  direction, $\vec{p}(B)$, and the direction of the decay vertex - estimated
  as the vector originating at the primary vertex and terminating at the 
  muon-pair vertex;
  
\item[$\mathbf{\iso}$:] the isolation of the candidate \bs\ defined as,
  $\iso=p_{T}^{\bs}/(p_{T}^{\bs}+\sum_i p_{T}^{i}(\Delta R<1.0))$,
  where the sum is over all tracks within an $\eta-\phi$ cone of radius
  $R=1.0$, centered on $\vec{p}(B)$;

\item[$\mathbf{|d_0(\mu_1)|}$:] impact parameter of muon which has the larger
  value of the muon pair;
  
\item[$\mathbf{|d_0(B^0_s)|}$:] impact parameter of $B^0_s$ reconstructed from
  muon pair;
  
\item[$\mathbf{L_{\mbox{\tiny 2d}}/\sigma_{L_{\mbox{\tiny 2d}}}}$:] significance of $L_{\mbox{\tiny 2d}}$;
  
\item [$\mathbf{\chi^2_{\mbox{\tiny vtx}}}$:] $\chi^2$ of the vertex;

\item[$\mathbf{L_{\mbox{\tiny 3D}}}$:] 3-dimensional vertexes displacement
  obtained by the vertex fitting;
  
\item[$\mathbf{p_T (\mu_2)}$:] the transverse momentum of lower momentum
  daughter muon;	
  
\item[$\mathbf{|d_0(\mu_2)|/\sigma_{d_0(\mu_2)}}$:] Significance of smaller $|d_0(\mu)|$
  of the muon with smaller impact
  parameter, where $\sigma_{d_0(\mu)}$ is estimated uncertainty of
  $|d_0(\mu)|$;
  
\item[$\mathbf{\lambda/\sigma_{\lambda}}$:] proper decay length significance 
  ($\lambda$ divided by the uncertainty on $\lambda$);

\item[$\mathbf{\lambda}$:] the proper decay length;
  
\item[$\mathbf{|d_0(\mu_1)|/\sigma_{d_0(\mu_1)}}$:] significance of larger $|d_0(\mu)|$
  of the muon with larger impact
  parameter, where $\sigma_{d_0(\mu)}$ is estimated uncertainty of
  $|d_0(\mu)|$;
  
\item[$\mathbf{\Delta\alpha_{xy}}$:] the 2D opening angle between the \bs\ flight 
  direction, $\vec{p}(B)$, and the direction of the decay vertex - estimated
  as the vector originating at the primary vertex and terminating at the 
  muon-pair vertex;
  
\item[$\mathbf{|d_0(\mu_2)|}$:] impact parameter of muon which has the smaller
  value of the muon pair.
\end{description}

The final NN network consists of 14 input variables.  Extensive cross checks were done to demonstrate the 14-variable network does not sculpt the di-muon invariant mass distribution. Figure~\ref{fig:backsculpt} shows the correlation across the \Mmm mass range.  No significant correlations are seen. 
\vspace*{0.10in}

% --- bmass vs NN cut

\subsection{Expected Limits for CDF}
A comparison of the previous expected and observed limits is shown in Table~\ref{tab:expLims}. 
The expected limit after including the new data set, the increased acceptance, and the NN improvements
is about \brbsmm$\sim2\times 10^{-8}$.
\begin{table}[!hbtp]
\begin{center}
\begin{tabular}{|l|cc|}
\hline & \textbf{Expected} & \textbf{Observed} \\ 
\hline $2 \textnormal{fb}^{-1}$ &  4.9$\times 10^{-8}$ & 5.8$\times 10^{-8}$\\
\hline $3.7 \textnormal{fb}^{-1}$ &  3.3$\times 10^{-8}$ & 4.3$\times 10^{-8}$\\
\hline $6.9 \textnormal{fb}^{-1}$ &  $\sim2\times 10^{-8}$ & -\\
\hline
\end{tabular}
\caption{CDF expected and observed limits.}
\label{tab:expLims}
\end{center}
\end{table}

\section{Conclusion}
\label{sec:Conclusion}
FCNC's are powerful probe for New Physics and CDF and \dzero\ continue to lead
the searches in the B sector. CDF is updating their \bsdmm\ analysis which will significantly 
improve the expected limit for \brbsdmm. CDF also has the first observation of $B^{0}_{s}\rightarrow \mu^+\mu^-\phi$
as well as a \afb\ measurement that competes and agrees with the B-factory results. Shortly after the FPCP conference 
CDF updated both the $b\to s\mu^{+}\mu^{-}$ and \bsdmm analysis \cite{bmmUpdate, btosmmAsym, lambdabtolambdamm}.

% =============================================================================

\end{document}

%% file: econfmacros.tex
\def\beq{\begin{equation}}
\def\eeq#1{\label{#1}\end{equation}}
\def\eeqn{\end{equation}}

\def\beqa{\begin{eqnarray}}
\def\eeqa#1{\label{#1}\end{eqnarray}}
\def\eeqan{\end{eqnarray}}

\let\bar=\overbar

\def\Dslash{\not{\hbox{\kern-4pt $D$}}}
\def\dslash{\not{\hbox{\kern-2pt $\del$}}}

\def\msb{{\bar{\ssstyle M \kern -1pt S}}}